\documentclass[12pt]{article}

\usepackage{epsf}
\usepackage[dvips]{graphicx}

\def\bk{{\mbox{\boldmath$k$}}}
\def\bq{{\mbox{\boldmath$q$}}}

\def\bgam{{\mbox{\boldmath$\gamma$}}}
\def\ds{\displaystyle}

\begin{document}

\begin{center}
{\bfseries SOLVING THE INHOMOGENEOUS BETHE-SALPETER EQUATION}

\vskip 5mm

S.S. Semikh$^{1 \dag}$, S.M. Dorkin$^{2}$, M. Beyer$^{3}$, L.P.
Kaptari$^{1}$

\vskip 5mm

{\small (1) {\it BLTP, Joint Institute for Nuclear Research }
\\
(2) {\it SNPI, Moscow State University }
\\
(3) {\it University of Rostock, Germany }
\\
$\dag$ {\it E-mail: semikh@thsun1.jinr.ru }}
\end{center}

\vskip 5mm

\begin{center}
\begin{minipage}{150mm}
  \centerline{\bf Abstract}
We develop an advanced method of solving homogeneous and inhomogeneous
Bethe-Salpeter equations by using the expansion over the complete set of
4-dimensional spherical harmonics. We solve Bethe-Salpeter equations for
bound and scattering states of scalar and spinor particles for the case of
one meson exchange kernels. Phase shifts calculated for the scalar model
are in agreement with the previously published results. We discuss
possible manifestations of separability for one meson exchange interaction
kernels.
\end{minipage}
\end{center}

\section{Introduction and preliminaries}

Obtaining solutions of inhomogeneous Bethe-Salpeter (BS) equations
is a long standing problem with clear motivation -- BS amplitude
in continuum (or half-off-shell $T$-matrix) is a necessary
ingredient for description of final state interaction in numerous
processes with the correlated nucleon-nucleon pair in final state.
As an example one can mention, for instance, charge-exchange
reaction on deuteron $pD\to n(pp)$ \cite{our_chex}, deuteron
break-up with fast forward $pp$-pair \cite{our_brkp}, electro- and
photo-disintegration of deuteron \cite{edenp} etc. One of the
possible ways to obtain the BS amplitude in continuum is the
so-called one iteration approximation scheme \cite{OIA}, when the
relativistic $P$-components are obtained as first iteration from
the non relativistic wave function with a certain interaction
kernel. In \cite{tjon} the inhomogeneous BS equation for spinor
particles is regularly treated with a  2-dimensional Gaussian
mesh. Although there are no doubts in correctness of these
results, it can be shown that, in general, inappropriate choice of
a 2-dimensional mapping in integration procedure can lead to
under- or overestimation of the solutions. In this sense a
2-dimensional change of variables that one has to apply in partial
BS equation for transforming the limits of integration to finite
ones is not well defined procedure. Thus, the search of the
rigorous method to solve such an equations should be continued.

In \cite{fb-tjon} the expansion of the BS amplitude and
interaction kernel over hyperspherical basis is employed to
formulate a method of solving the BS equation for bound states of
scalar particles,
\begin{eqnarray}
\label{sceq}
\Gamma(k)=i\int\frac{d^4p}{(2\pi)^4}\,V(k,p)\,S(p_1)S(p_2)\,\Gamma(p),
\end{eqnarray}
$p_{1,2}=P/2\pm p$, and $S(p)=(p^2-m^2+i\varepsilon)^{-1}$. The
corresponding expansions after the Wick rotation looks like
\begin{eqnarray}
\label{exp1}
\Gamma(ik_4,\bk)&=&\sum\limits_{nlm}\varphi_l^n(\tilde k)\,Z_{nlm}(\omega_k),\\
V(k,p)=\frac {g^2}{(k-p)^2+\mu^2}&=&2\pi^2 \sum_{nlm}
\frac{1}{n+1} V_n(\tilde k,\tilde p)Z_{nlm}(\omega_k)
Z_{nlm}^{*}(\omega_p),\label{exp2}\\ V_n(a,b)&=& \frac
{4g^2}{(\Lambda_+ +\Lambda_-)^2}
\left( \frac {\Lambda_+ -\Lambda_-}{\Lambda_+ +\Lambda_-}\right)^n,\nonumber \\
\Lambda_{\pm}&=&\sqrt{(a\pm b)^2+\mu^2},\nonumber
\end{eqnarray}
where by definition $\tilde k=\sqrt{k_4^2+\bk^2}$, and
$\omega_k=(\chi,\theta,\phi)$ -- angles of vector $k=(k_4,\bk)$ in
4-dimensional Euclidean space. Hyperspherical harmonics
\begin{eqnarray}
Z_{nlm}(\chi,\theta,\phi)=X_{nl} (\chi) Y_{lm}(\theta,\phi),\quad
X_{nl} (\chi)\sim \sin^l\chi C_{n-l}^{l+1}(\cos \chi),
\nonumber\end{eqnarray} are proportional to the product of usual
spherical harmonics for orbital momentum $l$ and Gegenbauer
polynomials $C_{n-l}^{l+1}$.

Performing a  Wick rotation in (\ref{sceq}) and substituting
(\ref{exp1}) and (\ref{exp2}) one reduces the BS equation to the
infinite set of 1-dimensional integral equations. Thus, there is
only one degree of freedom left in the choice of the mapping for
integration. It is shown in \cite{fb-tjon}, that this approach is
very effective for numerical purpose. In particular, due to rapid
convergence of hyperspherical expansion, the consideration of only
first few equations from the set is sufficient.

\section{Inhomogeneous BS equation for scalar particles}

In this paper the method introduced in \cite{fb-tjon} is applied
to the scalar inhomogeneous BS equation, and generalized to spinor
equations, both homogeneous and inhomogeneous. Let us first
consider the BS equation for the scattering of scalar particles:
\begin{eqnarray}
T(k)=V(k,q)+i\int\frac{d^4p}{(2\pi)^4}\,V(k,p)\,S(p_1)S(p_2)\,T(p).\label{ihhom}
\end{eqnarray}
Here $T(k)$ is half-off-shell $T$-matrix for the scattering of
real particles with momenta $q_{1,2}$, and in c.m.s., where the
real 3-momentum of particles $|\bq|=\hat p$,
\begin{eqnarray}
s=(q_1+q_2)^2=(2E_{\hat p})^2,\quad E_{\hat p}=\sqrt{m^2+{\hat
p}^2},\quad q=\frac{q_1-q_2}{2},\quad (2m)^2<s<(2m+\mu)^2\nonumber
\end{eqnarray}
Before Gegenbauer decomposition of type (\ref{exp1}) and
(\ref{exp2}) the Wick rotation should be applied. It is more
convenient to perform firstly in (\ref{ihhom}) the usual partial
expansion over the spherical harmonics $Y_{lm}$ (see also
\cite{nash_yaf1}):
\begin{eqnarray}
T_l(k_0,k)=g^2Q_l(k_0,k;0,{\hat
p})-i\int\frac{dp_0\,dp}{(2\pi)^3}\,g^2Q_l(k_0,k;p_0,p)\,S(p_1)S(p_2)\,T_l(p_0,p).
\label{part}
\end{eqnarray}
Here $Q_l(k_0,k;p_0,p)\equiv
Q_l([k^2+p^2+\mu^2-(k_0-p_0)^2-i\varepsilon]/2kp)$ is the adjoint
Legendre function of the 2nd kind. The normalization of $T$-matrix
is fixed by the free term in (\ref{part}) and leads to the following
expression for the phase shifts:
\begin{eqnarray} T_l(0,{\hat p})=16\,\pi\,{\hat
p}\,\sqrt{s}\,e^{i\delta_l}\,\sin\delta_l.\label{shift}
\end{eqnarray}
To eliminate the removable singularity at $(p_0,p)=(0,\hat p)$, we
present the $T$-matrix in the following factorized form
\cite{pr-tjon}:
\begin{eqnarray}
T_l(k_0,k)\sim \varphi_l(k_0,k)\,t_l(s),\quad t_l(s)\sim
e^{i\delta_l}\,\sin\delta_l,\nonumber
\end{eqnarray}
thus obtaining from (\ref{part}) the equation for $\varphi_l$,
\begin{eqnarray}
\nonumber \varphi_l(k_0,k)&=&g^2Q_l(k_0,k;0,{\hat p})
-i\int\frac{dp_0\,dp}{(2\pi)^3}\,\{g^2Q_l(k_0,k;p_0,p)\\
&-& \frac{g^2}{Q_l(s)}\,Q_l(k_0,k;0,{\hat p})\,Q_l(0,{\hat
p};p_0,p)\}\,S(p_1)S(p_2)\,\varphi_l(p_0,p),\label{part1}
\end{eqnarray}
$Q_l(s)\equiv Q_l(0,\hat p;0,\hat p)$. In the following only the
cases $l=0,1$ will be considered. After the Wick rotation the
Gegenbauer decomposition of $\varphi_l$ for $l=0$ takes the form
\begin{eqnarray}
\varphi_0(ik_4,k)=k\hat p\sum\limits_{j=1}^\infty g_j(\tilde k
)\,X_{2j-2,0}(\chi),\label{pamp}
\end{eqnarray}
and for the coefficient functions $g_j$ the system of integral
equations is obtained:
\begin{eqnarray}
g_j(\tilde k)&=&\frac{\pi}{2j-1}\,V_{2j-2}(\tilde k,\hat
p)\,X_{2j-2,0}\left(\frac{\pi}{2}\right)\nonumber\\\nonumber
&+&\sum\limits_{l=1}^\infty \int\limits_{0}^\infty\frac{d\tilde
p\, {\tilde p}^3}{8\pi^2}\left[ \frac{1}{2j-1}\,V_{2j-2}(\tilde
k,\tilde p)\, S^0_{2j-2,2l-2}(\tilde p)\right.\\\nonumber
&-&\left.\frac{{\hat p}^2}{2j-1}\,V_{2j-2}(\tilde k,\hat
p)\,X_{2j-2,0}\left(\frac{\pi}{2}\right)\frac{1}{Q_0(s)}\,N_{0,2l-2}(\hat
p,\tilde p)\right]g_l(\tilde p)\\\nonumber
&-&\int\limits_{0}^{\hat
p}\frac{dp}{8\pi^2}\,\frac{1}{E_p\sqrt{s}\,(\sqrt{s}-2E_p)}\left[\frac{p}{\hat
p}\,g^2\,W_{2j-2,0}(\tilde k,p)\right.\\\label{gg1}
&-&\left.\frac{2}{Q_0(s)}\,\frac{\pi}{2j-1}\,V_{2j-2}(\tilde
k,\hat p)\,X_{2j-2,0}\left(\frac{\pi}{2}\right)\,Q_0(0,\hat
p;p_2^0,p)\right]\tau_0(p),\\[5mm]\nonumber
\tau_0(k)&=&g^2\,Q_0(k_2^0,k;0,\hat p)+g^2\hat
p\sum\limits_{l=1}^\infty \int\limits_{0}^\infty\frac{d\tilde p\,
{\tilde p}^3}{(2\pi)^3}\,\left[k\,N_{0,2l-2}(k,\tilde
p)\right.\\\nonumber &-&\left.\frac{\hat p
}{Q_0(s)}\,Q_0(k_2^0,k;0,\hat p)\,N_{0,2l-2}(\hat p,\tilde
p)\right]g_l(\tilde p)\\\nonumber &-&g^2\int\limits_{0}^{\hat
p}\frac{dp}{8\pi^2}\,\frac{1}{E_p\sqrt{s}\,(\sqrt{s}-2E_p)}\left[
k\,p\,U_0(k,p)\right.\\\label{gg2}
&-&\left.\frac{2}{Q_0(s)}\,Q_0(k_2^0,k;0,\hat p)\,Q_0(0,\hat
p;p_2^0,p)\right]\tau_0(p).
\end{eqnarray}
The additional unknown function $\tau_0$ comes after the Wick
rotation from the residue at the pole
$p_2^0=\sqrt{s}/2-E_p+i\varepsilon$, which is able to cross the
imaginary $p_0$ axis. It is proportional to the $T$-matrix in
Minkowsky space for certain value of relative energy,
$\tau_0(p)\sim T_0(\sqrt s/2-E_p,p)$. The partial kernels are
\begin{eqnarray}
S^l_{k^{\prime}k}(\tilde p)&=&\int \limits_0^{\pi} d\chi \sin^2
\chi\, \frac {X_{k^{\prime}l}(\chi)X_{kl}(\chi)}{({\tilde
p}^2-{\hat
p}^2)^2+s{\tilde p}^2\cos^2\chi},\nonumber\\
N_{ml}(k,\tilde p)&=&\int \limits_0^{\pi} d\chi_p \sin^2 \chi_p\,
\frac{1}{2kp}\{Q_m(k_2^0,k;ip_4,p)\nonumber\\
&+&Q_m(-k_2^0,k;ip_4,p)\}\frac {X_{lm}(\chi_p)}{({\tilde
p}^2-{\hat
p}^2)^2+s{\tilde p}^2\cos^2\chi_p},\nonumber\\
W_{nl}(\tilde k,p)&=&\int \limits_0^{\pi} d\chi_k \sin^2 \chi_k\,
X_{nl}(\chi_k)\,\frac{1}{kp}\{Q_l(ik_4,k;p_2^0,p)
+Q_l(ik_4,k;-p_2^0,p)\},\nonumber\\
U_l(k,p)&=&\frac{1}{kp}\{Q_l(k_2^0,k;p_2^0,p)
+Q_l(k_2^0,k;-p_2^0,p)\}.\nonumber
\end{eqnarray}
It is easy to see that all the partial kernels are real
expressions.

\section{Numerical results and separability}

As all the integrations in system (\ref{gg1}), (\ref{gg2}) are
1-dimensional, the integrals can be replaced by finite sums by
using the gaussian mesh. In this way this system is reduced to
the usual system of linear equations, which can be solved by any
appropriate method. Besides, the solution of integral equations
(\ref{gg1}), (\ref{gg2}) can be obtained by iterating the free
term, thus constructing the Neumann series, which represents the
solution of an inhomogeneous integral equation. It was explicitly
checked that these two procedures of finding a solution lead to
the same results.

\begin{figure}[ht]
~\centering \epsfxsize 4.5in \epsfbox{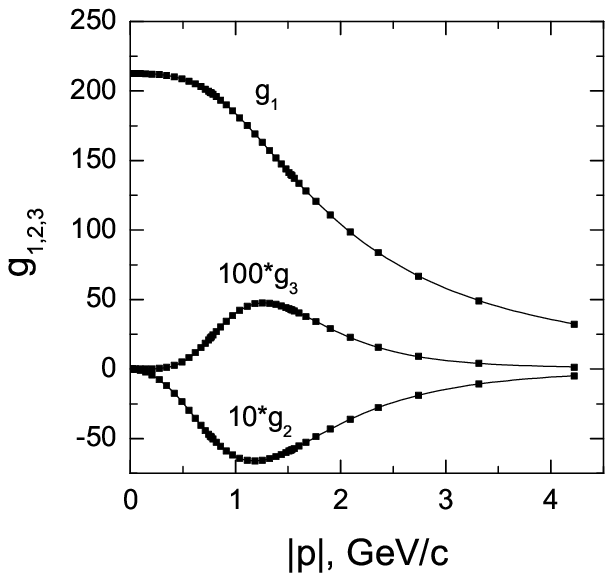} ~\vskip -6mm
\caption{Gegenbauer components from (\protect\ref{pamp}).
Functions $g_2$ and $g_3$ are multiplied by 10 and 100
respectively in view of their smallness.}\label{pict1}
\end{figure}

Taking the first five terms in the expansion (\ref{pamp}), the
system (\ref{gg1}), (\ref{gg2}) was numerically considered for the
following set of parameters:
\begin{center}
$\ds{\frac{g^2}{4\pi}=4\pi}$, $m=\mu=1$ GeV, ${\hat p}=0.77$ GeV/c
\end{center}
In Fig. \ref{pict1} the numerical results for the first three
coefficient functions from (\ref{pamp}) are shown. It was
established that convergence of this expansion is quite rapid,
that is why only 3 components are presented and $g_2$ and $g_3$
are multiplied by 10 and 100 respectively. Centered symbols
correspond to the points of the mesh, where the solution is
defined. Lines connecting the points reproduce the results of the
fit for the obtained Gegenbauer components with the fitting
functions\\[3mm] 1. For $g_1(p),
p\equiv \tilde p$
\begin{eqnarray}
\label{fit1}
F(p)=\sum\limits_{j=1}^4\frac{a_j^1\,p^{2j-2}}{(p^2+b_1^2)^j},
\end{eqnarray}
2. For $g_2(p)$
\begin{eqnarray}
\label{fit2}
F(p)=\frac{p^2}{p^2+b_2^2}\sum\limits_{j=1}^4\frac{a_j^2\,p^{2j-2}}{(p^2+b_2^2)^j},
\end{eqnarray}
3. For $g_3(p)$
\begin{eqnarray}
\label{fit3} F(p)=\left[\frac{p^2}{p^2+b_3^2}\right]^2
\sum\limits_{j=1}^4\frac{a_j^3\,p^{2j-2}}{(p^2+b_3^2)^j}
\end{eqnarray}
For the sake of brevity we don't show here the numerical values of
the parameters $a_j^{1,2,3}$ and $b_{1,2,3}$, which can be adjusted by
any appropriate method (for instance, the built-in fitting
procedure in Microcal Origin 7.0 can be employed). It should be
mentioned that the quality of this analytical fit is excellent in
the whole range of the argument $\tilde p$ despite the very simple
form of the functions (\ref{fit1})-(\ref{fit3}). This fact can be
treated as an indication to some kind of separability in the one meson
exchange interaction and will be discussed below. Besides, it was
found that such a fit is valid for Gegenbauer components of
solutions of any BS equation in ladder approximation (scalar and
spinor equations, both homogeneous and inhomogeneous). Hence, it
is very useful for practical purposes, e.g. to represent the
numerical results in a compact form.

To discuss the manifestations of separability in more details, let
us now turn to the homogeneous BS equation (\ref{sceq}). After
Wick rotation and usual partial decomposition it gives
\begin{eqnarray}
\Gamma_l(ik_4,k)=\int\frac{dp_4\,dp\,p^2}{(2\pi)^3}\,
V_l(ik_4,k;ip_4,p)\,S(p_1)S(p_2)\,\Gamma_l(ip_4,p). \label{hpart}
\end{eqnarray}
It is generally known that for the conventional separable kernel
of the form (see e.g. \cite{burov})
\begin{eqnarray}
V_l(ik_4,k;ip_4,p)=\sum \limits_{ij} \lambda_{ij}(s)\, g_i({\tilde
k}^2)\, g_j({\tilde p}^2)\nonumber
\end{eqnarray}
the vertex functions are expressed in terms of functions $g$:
\begin{eqnarray}
\Gamma_l(ik_4,k)= \sum\limits_{i}c_i\,g_i({\tilde k}^2). \nonumber
\end{eqnarray}
In our case, solving equation (\ref{hpart}) for kernel
(\ref{exp2}) by the method given above and fitting the calculated
Gegenbauer components with analytical expressions like
(\ref{fit1})-(\ref{fit3}), we obtain the following approximate
representation for the solution (cf. (\ref{pamp})):
\begin{eqnarray}
\Gamma_l(ik_4,k)=\sum\limits_{j=l}^N c_j^l\,R_{jl}({\tilde k}^2)\,
X_{jl}(\chi_k),\label{sep1}
\end{eqnarray}
where the fitting functions are
\begin{eqnarray}
R_{jl}({\tilde k}^2)=\frac{({\tilde k}^2)^{\mu_{jl}}}{({\tilde
k}^2+\beta_{jl}^2)^{{\nu_{jl}}}}.\nonumber
\end{eqnarray}
It is clear that constructing the expression for the separable
kernel like
\begin{eqnarray}
V_l(ik_4,k;ip_4,p)=\sum\limits_{ij} \lambda_{ij}(s)\,
g_{il}(p_4,p)\, g_{jl}(k_4,k)\label{sepkern1}
\end{eqnarray}
with
\begin{eqnarray}
g_{jl}(k_4,k)=R_{jl}({\tilde k}^2)\,
X_{jl}(\chi_k)\label{sepkern2}
\end{eqnarray}
one identically reproduces the analytical form of the solution
(\ref{sep1}). Therefore, the kernel (\ref{sepkern1}),
(\ref{sepkern2}) can be referred to as some general form for
constructing separable kernels. Note that it has the following
properties:
\begin{itemize}
\item A dependence on mass (or $s$) in the functions $g$
\item An explicit dependence on $k_4$ in $g$ coming from Gegenbauer
polynomials being the functions of $\cos\chi_k=k_4/\tilde k$
\item At low $k^2$ (\ref{sepkern1}) tends to a Yamaguchi kernel, because
$g(k^2)\to \ds{\frac {C}{k^2+\beta^2}}$
\end{itemize}
A concrete example of the solution of eq. (\ref{hpart}) is given in
Table 1. The parameters $\lambda_{ij}$ [GeV$^5$] for the partial kernel
(\ref{sepkern1}) of rank 3 are
\begin{eqnarray}
\lambda_{11}=0.19613  \quad \lambda_{22}= 0.13734  \quad
\lambda_{33}=-0.01644\nonumber\\
\lambda_{12}=-0.09675  \quad \lambda_{13}=0.00236 \quad
\lambda_{23}=-0.00034.\nonumber
\end{eqnarray}
\begin{table}[h]
\[
\begin{array}{ccccc}
\hline\hline j & c_j^0 &\beta_{j0}[\mbox{GeV}]&\mu_{j0}&\nu_{j0}
\\
\hline
0& 0.17904 & 0.39736 & 0 & 1 \\[1ex]
2& -0.00923& 0.41533 & 1 & 2 \\[1ex]
4& 0.00127& 0.43362 & 2 & 3 \\[1ex]
6& -0.00055 & 0.45127 & 3 & 4 \\[1ex]
8& 0.00017 & 0.47019 & 4 & 5 \\[1ex]
 \hline\hline
\end{array}
\]
\end{table}

\begin{minipage}{140mm}
Table 1. Numerical values of parameters in (\ref{sep1}) for $l=0$,
$m=1$ GeV, mass of bound state $M=1.9$ GeV, $\mu=0.1$ GeV.
Coefficients $c_j^0$ for odd values of $j$ are equal to zero.
\end{minipage}

\vskip 5mm

\begin{figure}[ht]
~\centering \epsfysize 4.5in \epsfbox{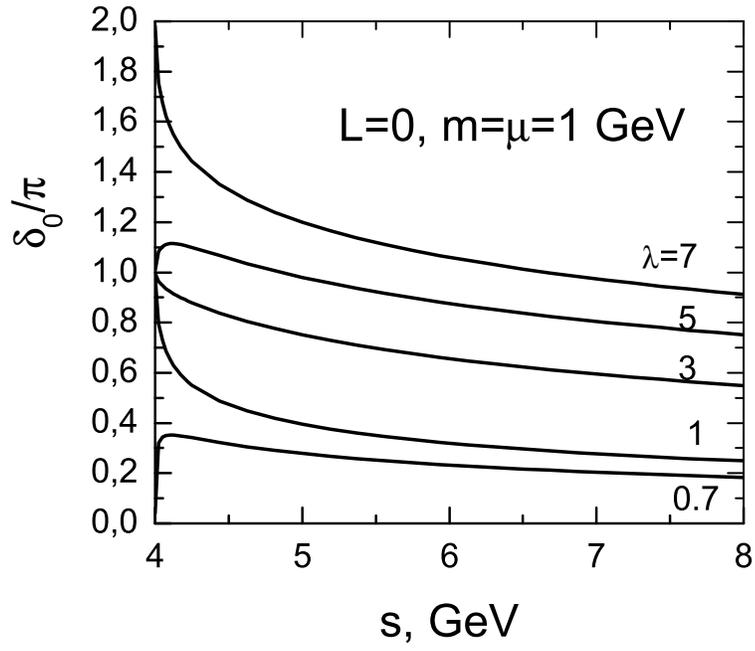} \caption{Phase
shifts (in the units of $\pi$) for $l=0$ and different values of
coupling constant.}\label{pict2}
\end{figure}

\begin{figure}[ht]
~\centering \epsfysize 4.5in \epsfbox{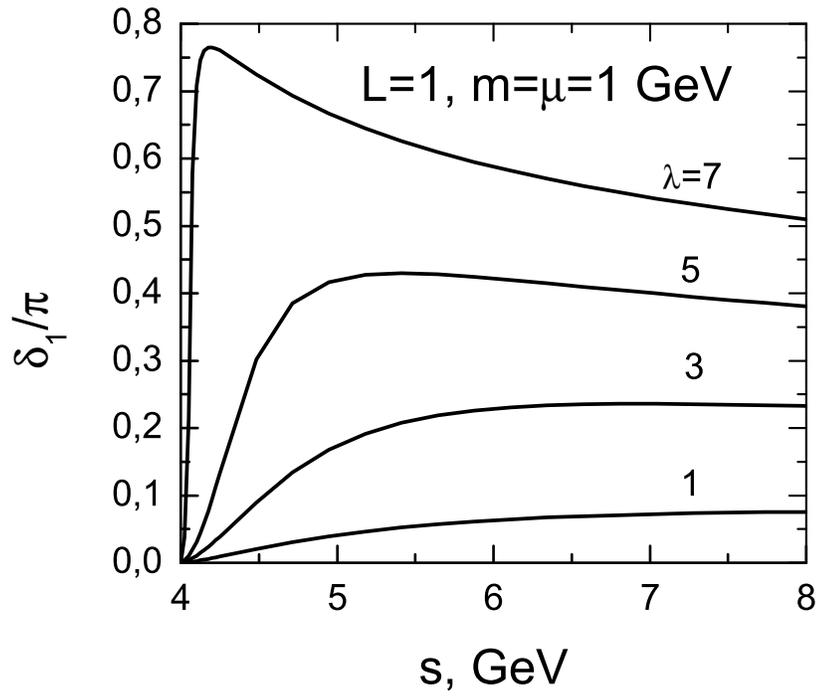} \caption{The
same as in Fig. \protect\ref{pict2} for $l=1$.}\label{pict3}
\end{figure}

\noindent Analyzing the obtained solutions for the system
(\ref{gg1}), (\ref{gg2}) it is also informative to calculate
corresponding phase shifts. The results for $l=0$ and $l=1$ are
presented in Figs. \ref{pict2} and \ref{pict3}, respectively, for
different values of coupling constant $\lambda$ connected with
$g^2$ as
\begin{eqnarray}
\nonumber \frac{g^2}{4\pi}=4\pi\lambda.
\end{eqnarray}
It is obvious, that for $\lambda=0.7$ there are no bound states in
this system due to the Levinson's theorem. For the values
$\lambda=1,3,5$ there is one bound state, and two bound states for
$\lambda=7$. For $l=1$ no bound states found for these values of
$\lambda$, as it follows from Fig. \ref{pict3}. Obtained results
for the phase shifts are in a good agreement with \cite{pr-tjon}.

\section{Spinor BS equations}

Above the procedure of solving the BS equation and handling with
its solutions is described for the case of scalar particles.
Spinor equations are much more complicated, and so we do not show
here the detailed formulas like (\ref{gg1}), (\ref{gg2}).
Nevertheless, it should be stressed that all the basic steps
remain the same. Below we briefly mention only the main results
for spinor case.

The problem of obtaining the BS amplitude in the continuum (or
half-off-shell $T$-matrix) for spinor particles as a solution of
the inhomogeneous BS equation with realistic interaction kernel seems
to be of great importance. We present the results of solving the
BS equation in the $^1S_0$ channel
\begin{eqnarray}
{\hat T}(k)=V(k,q)\,\Gamma(1)\,\Gamma_{^1S_0}^{++}(\bq)\,\Gamma(2)
+i\int\frac{d^4p}{(2\pi)^4}\,V(k,p)\,\Gamma(1)\, {\hat
S}(p_1)\,{\hat T}(p)\,{\tilde S(p_2)}\,\Gamma(2)\label{spinhom}
\end{eqnarray}
in the ladder approximation for the cases of scalar
($\Gamma(1,2)=1$) and pseudoscalar ($\Gamma(1,2)=\gamma_5$) one
meson exchanges without cut-off. Here
\begin{eqnarray}
\nonumber {\hat S}(p)=\frac{\hat p +m}{p^2-m^2+i\varepsilon},\quad
{\tilde S}(p)=\frac{\hat p -m}{p^2-m^2+i\varepsilon}.
\end{eqnarray}
In (\ref{spinhom}) the BS vertex function ${\hat T}(k)$ being the
matrix $4\times 4$ before Gegenbauer decomposition of type
(\ref{exp1}) and (\ref{exp2}) should be expanded over some full
set of matrices. In $^1S_0$ channel there are only 4 such a
matrices usually called spin-angular harmonics (for their explicit
form see \cite{static}),
\begin{eqnarray}
\Gamma_{^1S_0^{++}}(-\bk),\quad \Gamma_{^1S_0^{--}}(-\bk),\quad
\Gamma_{^3P_1^e}(-\bk),\quad \Gamma_{^3P_1^o}(-\bk),\label{harms}
\end{eqnarray}
and corresponding expansion after the Wick rotation reads like
\begin{eqnarray}
{\hat T}(ik_4,\bk)=\sum\limits_\alpha t_\alpha(ik_4,|\bk|)
\,\Gamma_\alpha(-\bk)\label{spex}.
\end{eqnarray}
Unfortunately, after the expansion (\ref{spex}) the Gegenbauer
decomposition of resulting partial equations for
$t_\alpha(ik_4,|\bk|)$ can not be performed in a simple analytical
way. To make it possible, another set of spin-angular matrices for
$^1S_0$ channel was found:
\begin{eqnarray}
{\cal T}_1(\bk)=\frac{\gamma_5}{2},\quad{\cal
T}_2(\bk)=\frac{\gamma_0\gamma_5}{2}, \quad{\cal
T}_3(\bk)=-\frac{(\bk,\bgam)}{2|\bk|}\gamma_0\gamma_5, \quad{\cal
T}_4(\bk)=-\frac{(\bk,\bgam)}{2|\bk|}\gamma_5,\label{nharms}
\end{eqnarray}
and, similarly to (\ref{spex}),
\begin{eqnarray}
{\hat T}(ik_4,\bk)=\sum\limits_{j=1}^4g_j(ik_4,|\bk|) \,{\cal
T}_j(\bk)\nonumber.
\end{eqnarray}
It can be easily shown that there is a non-degenerate transformation
from set (\ref{nharms}) to (\ref{harms}). The point of using these
two bases in the $^1S_0$ subspace is that the set (\ref{nharms})
allows us a very simple Gegenbauer decomposition of partial equations
for $g_j(ik_4,|\bk|)$ in an analytical form, while the set
(\ref{harms}) is preferable in calculating the matrix elements of
the real processes \cite{our_chex,our_brkp}, because of its clear
physical meaning.

\begin{figure}[ht]
~\centering \epsfysize 4.5in \epsfbox{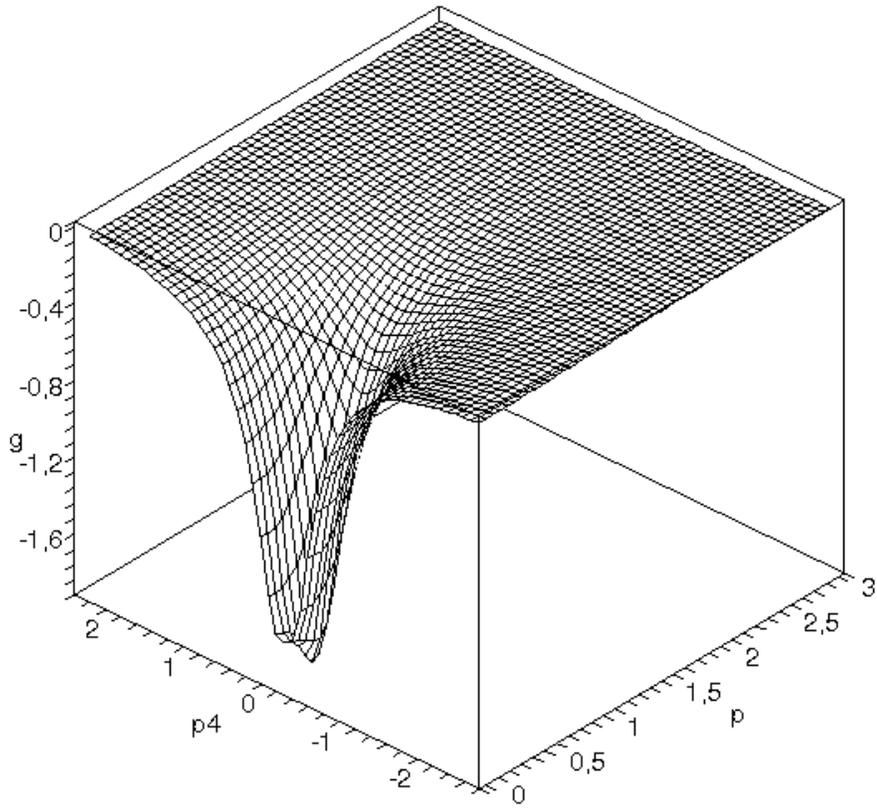}
\caption{$^1S_0^{++}$ partial component (the largest one) of the
solution of eq. (\protect\ref{spinhom}) for scalar meson exchange
interaction.}\label{pict4}
\end{figure}

\begin{figure}[ht]
~\centering \epsfysize 4.5in \epsfbox{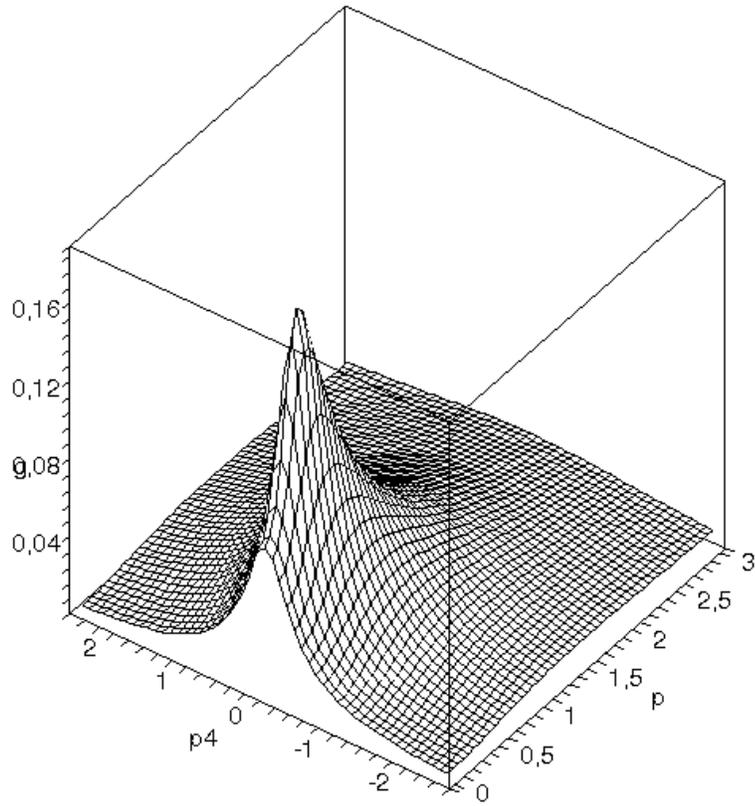}
\caption{$^1S_0^{--}$ partial component for scalar meson
exchange.}\label{pict5}
\end{figure}

\begin{figure}[ht]
~\centering \epsfysize 4.5in \epsfbox{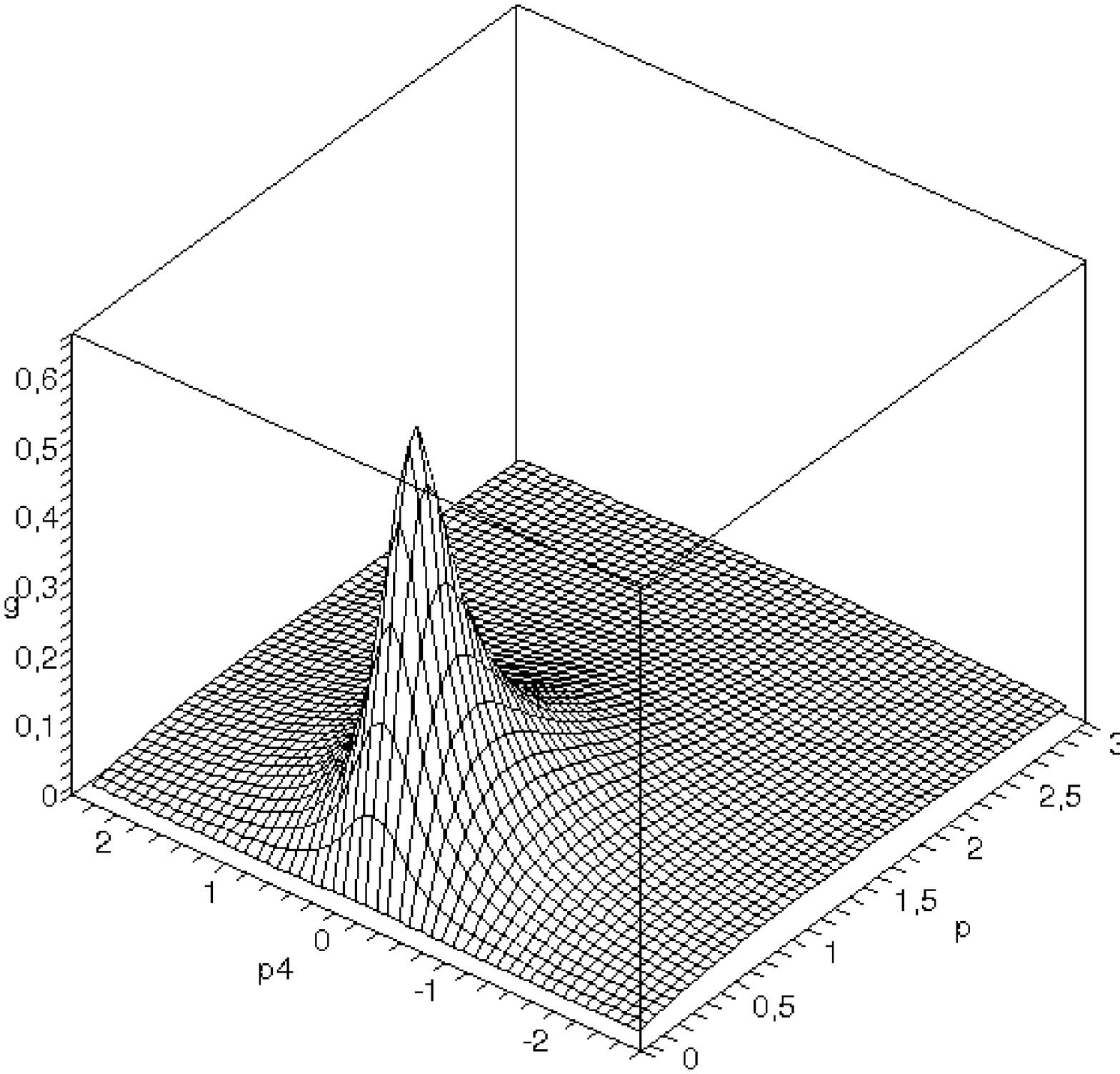} \caption{$^3P_1^e$
partial component for scalar meson exchange.}\label{pict6}
\end{figure}

\begin{figure}[ht]
~\centering \epsfysize 4.5in \epsfbox{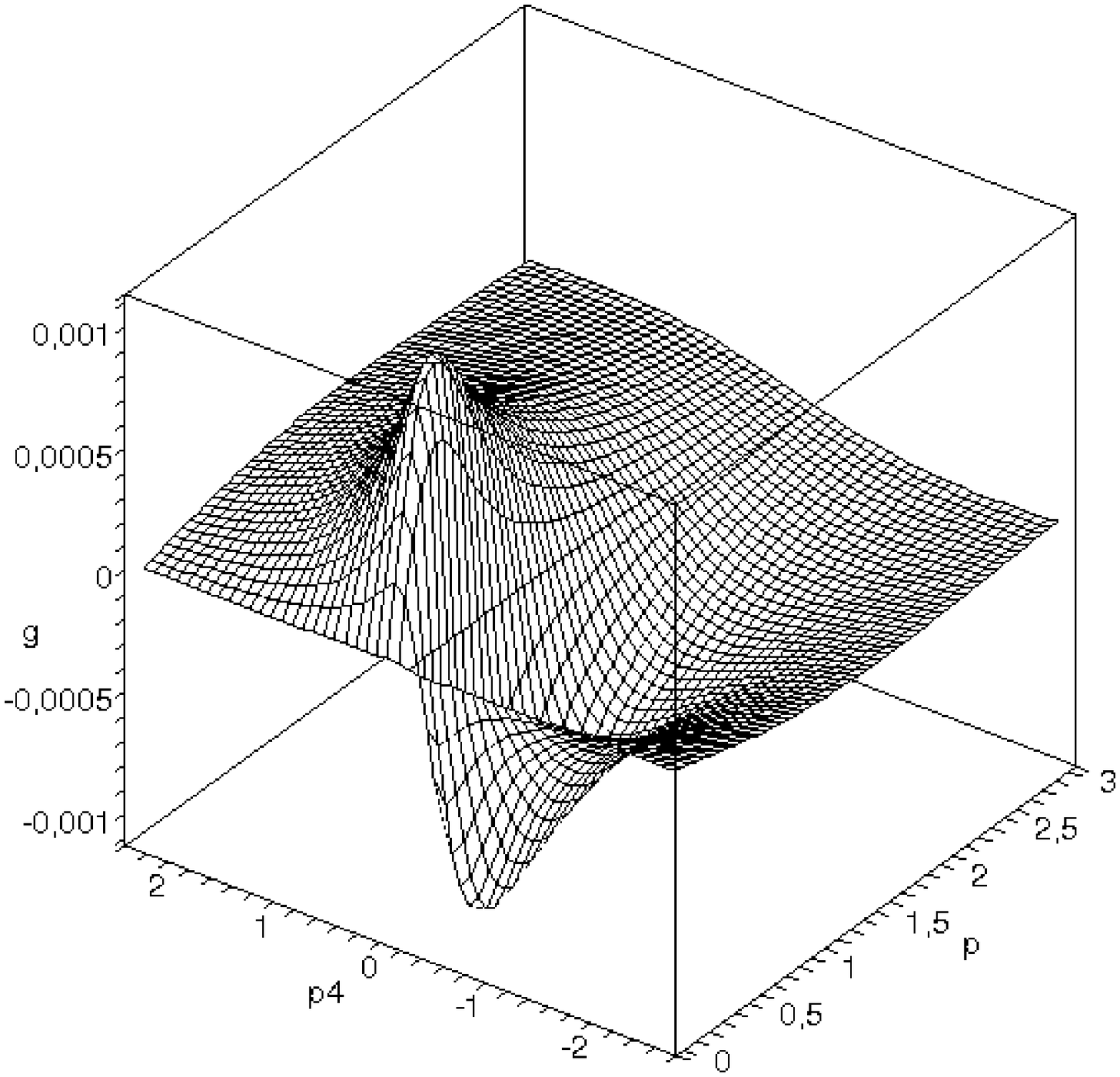} \caption{$^3P_1^o$
partial component for scalar meson exchange.}\label{pict7}
\end{figure}

\begin{figure}[ht]
~\centering \epsfysize 4.5in \epsfbox{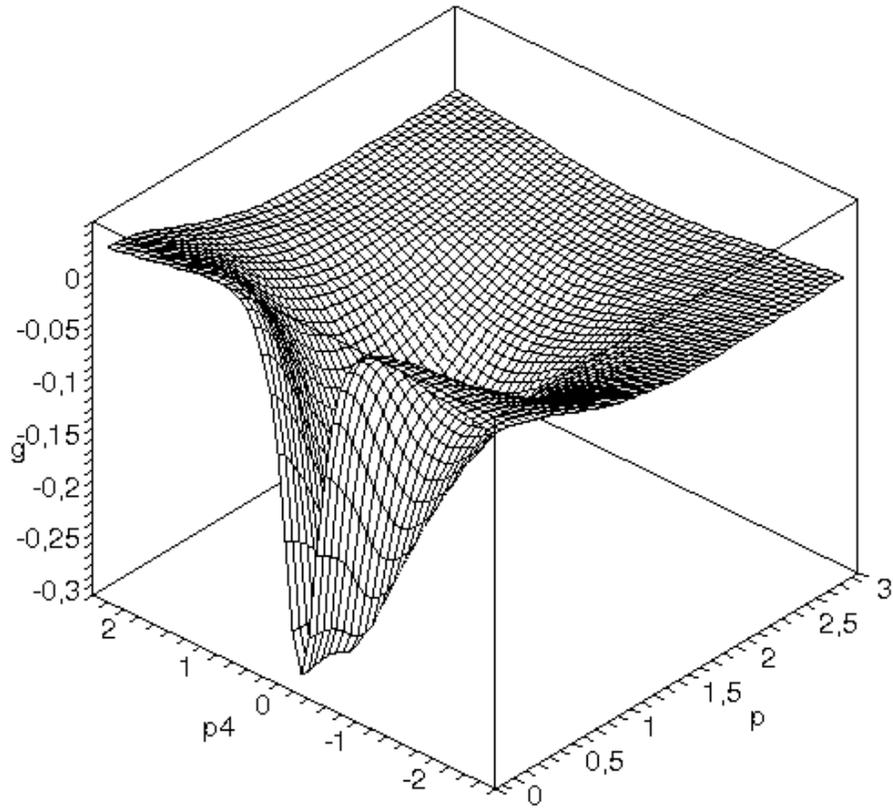}
\caption{$^1S_0^{++}$ partial component of the solution of eq.
(\protect\ref{spinhom}) for pseudoscalar meson exchange
interaction.}\label{pict8}
\end{figure}

\begin{figure}[ht]
~\centering \epsfysize 4.5in \epsfbox{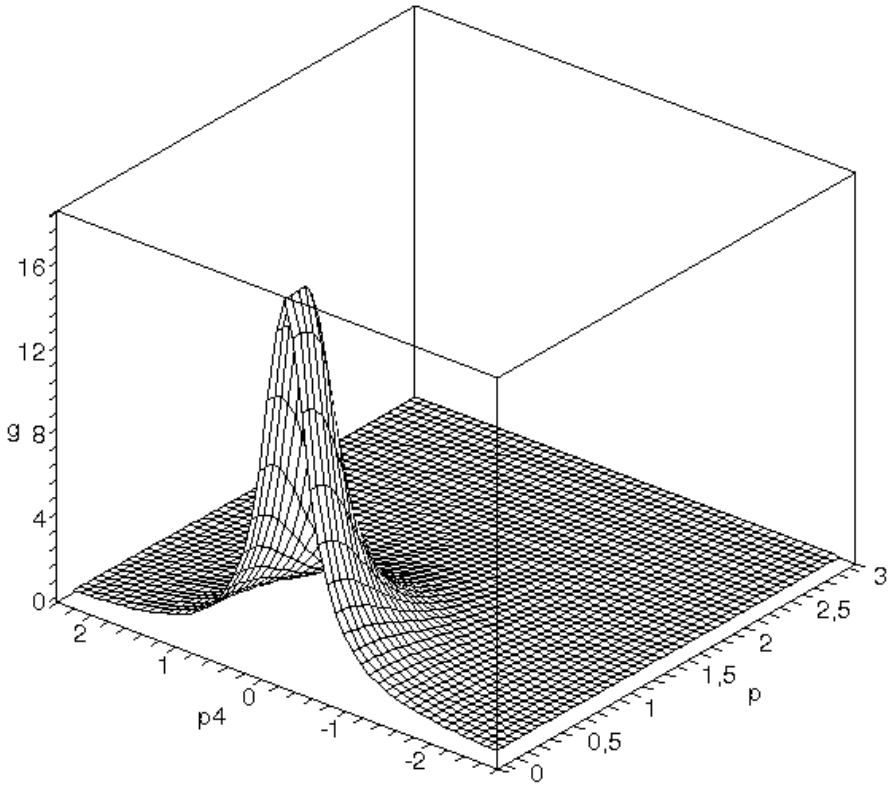}
\caption{$^1S_0^{--}$ partial component (the largest one) for
pseudoscalar meson exchange.}\label{pict9}
\end{figure}

\begin{figure}[ht]
~\centering \epsfysize 4.5in \epsfbox{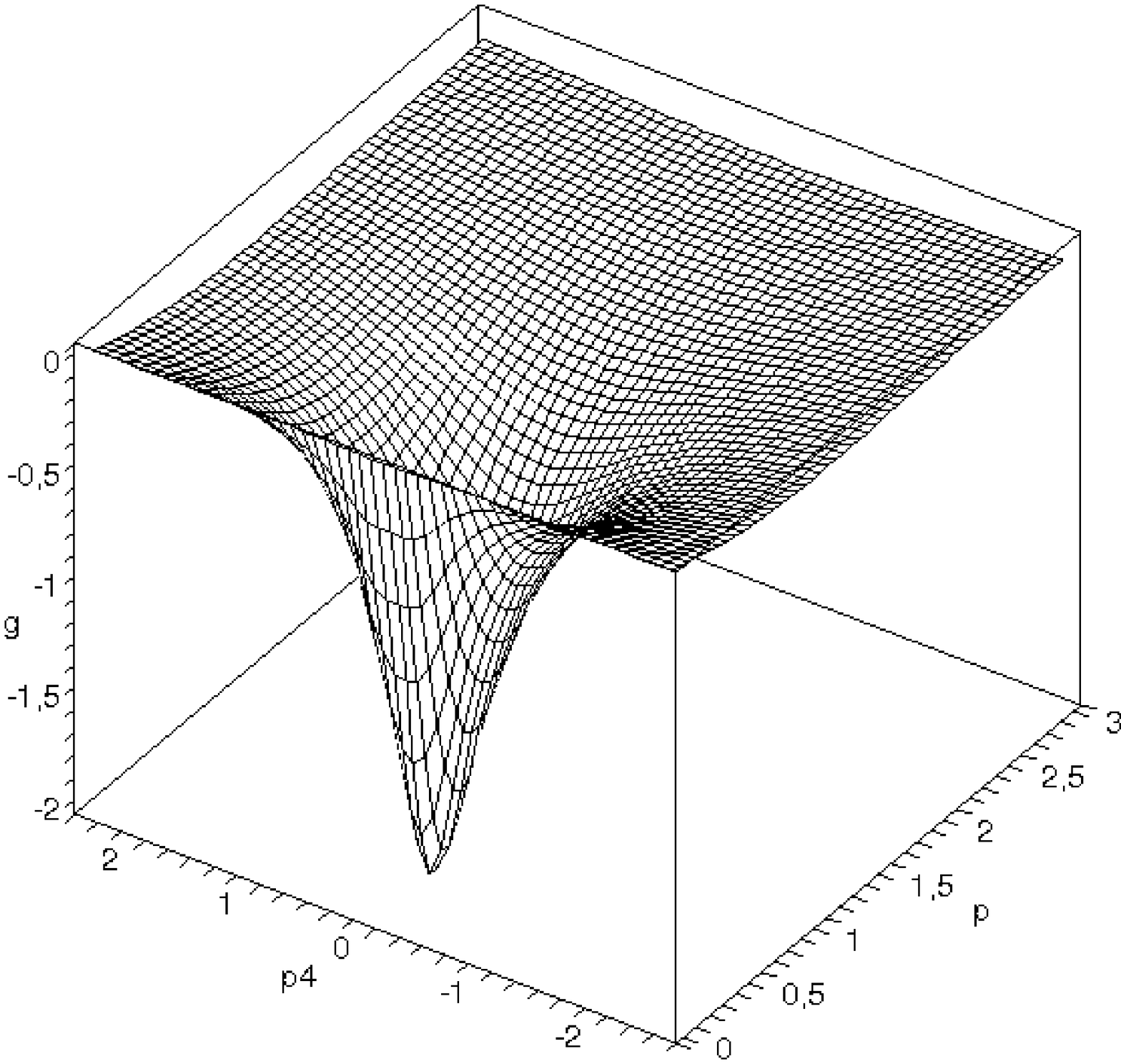} \caption{$^3P_1^e$
partial component for pseudoscalar meson exchange.}\label{pict10}
\end{figure}

\begin{figure}[ht]
~\centering \epsfysize 4.5in \epsfbox{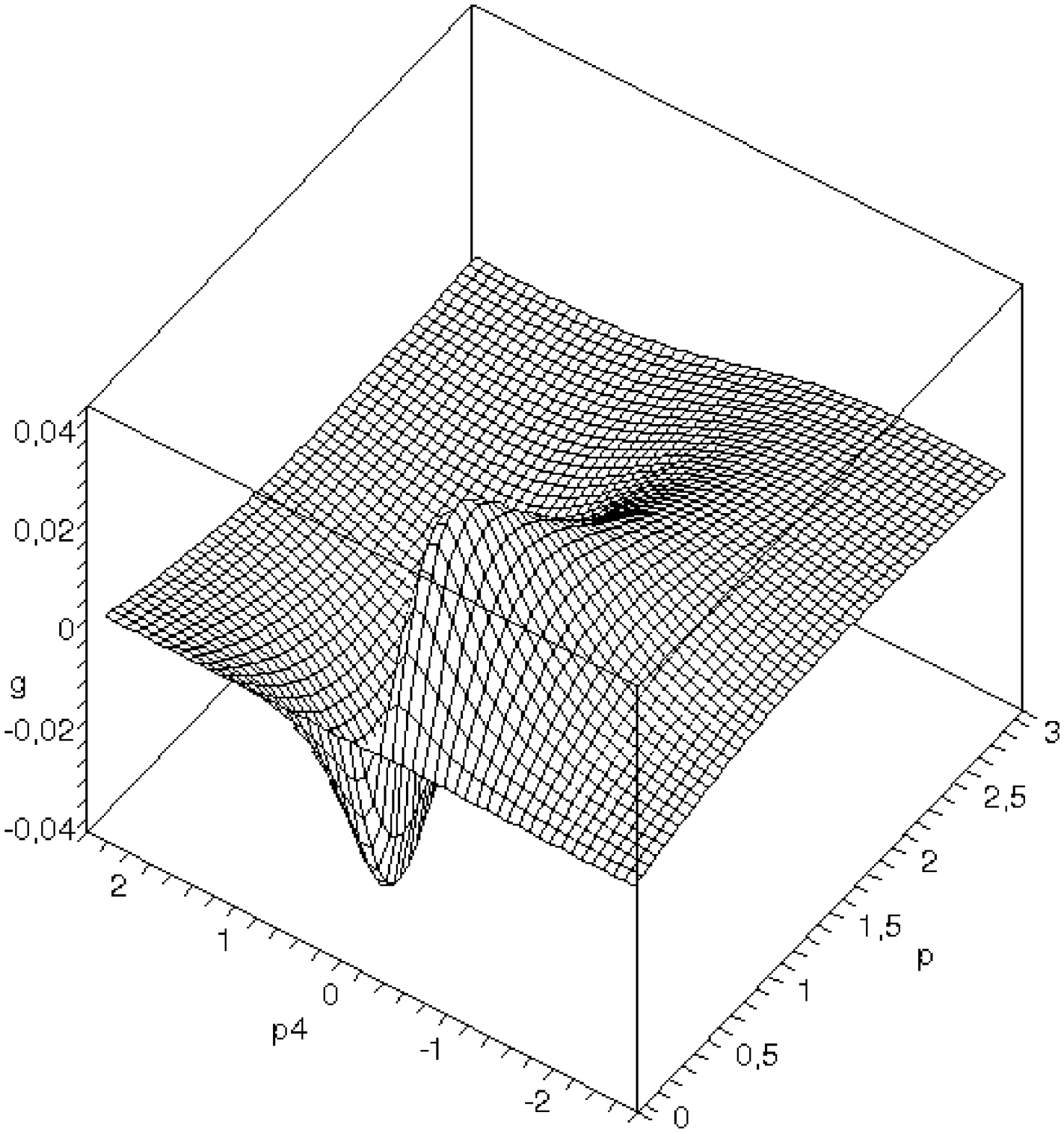} \caption{$^3P_1^o$
partial component for pseudoscalar meson exchange.}\label{pict11}
\end{figure}

From the statement above follows that the half-off-shell $T$-matrix ${\hat
T}(k)$ is expanded over the basis (\ref{nharms}) and the derived set
of integral equations for the scalar coefficients after the Wick
rotation is decomposed into Gegenbauer polynomials like in
(\ref{pamp})-(\ref{gg2}). We solve this set of equations
numerically, for $m=0.938$ GeV, $\mu=0.14$ GeV, $\sqrt{s}=2m+0.1$
GeV, and here the final results for solutions are presented in the
form of partial components $t_\alpha(ik_4,|\bk|)$ of
(\ref{spex}). In Figs. \ref{pict4}-\ref{pict7} these functions are
shown for the case of a scalar interaction, and in Figs.
\ref{pict8}-\ref{pict11} for a pseudoscalar one.  For scalar
exchange the ratio between different components looks very
familiar: $t_{^1S_0^{++}}$ is the largest one, the other components
are much smaller. Note that this situation does not
hold for pseudoscalar exchange: here $t_{^1S_0^{--}}$ becomes the
main component, and the $P$-wave $t_{^3P_1^e}$ is larger than
$t_{^1S_0^{++}}$. Although such a ratio between components seems
to be exotic, it is valid also for the bound state problem in the $^1S_0$
channel (eq. (\ref{sphom}) with $\Gamma(1,2)=\gamma_5$). Moreover,
it qualitatively reproduces the previous results (huge $P$-wave)
for the BS amplitude in continuum obtained in \cite{our_brkp}
within the one iteration approximation scheme.

Finally, the homogeneous spinor BS equation of the form
\begin{eqnarray}
\label{sphom}
\Psi(k)=i\int\frac{d^4p}{(2\pi)^4}\,V(k,p)\,\Gamma(1)\, {\hat
S}(p_1)\,\Psi(p)\,{\tilde S(p_2)}\,\Gamma(2),
\end{eqnarray}
was examined for one meson exchange interaction kernels with
$\Gamma(i), i=1,2$ as scalar, pseudoscalar, vector, tensor
vertices. Only $^1S_0$ bound states were considered. In this study
we followed the motivations of \cite{karmanov}, and the obtained
results will be a subject of a separate publication.

\section{Conclusion}

The method described before was successfully applied to solve both,
homogeneous and non homogeneous, BS equations for scalar and spinor
particles. It shows a high efficiency and accuracy, and, in
particular, does not require any extraordinary computer facilities.
The obtained results are in a good agreement with calculations of
other groups \cite{pr-tjon} and also agree with our previous
results \cite{our_brkp}. Also the manifestations of separability
for one meson exchange kernel are discussed. From the analysis of
the obtained exact solutions,  a possible general form of the
separable kernel is suggested.

\section{Acknowledgments}

We are grateful to K.Yu. Kazakov for valuable discussions. S.M.D.
and S.S.S. acknowledge the warm hospitality of the Elementary
particle physics group of the University of
Rostock, where part of this work was performed. This work is
supported by the Heisenberg - Landau program of JINR - FRG
collaboration and by the Deutscher Akademischer Austauschdienst.

\newpage

\newpage

\newpage

\newpage

\newpage

\newpage

\newpage

\newpage

\newpage

\newpage

\newpage

\end{document}